\documentclass{article}

\usepackage[utf8]{inputenc}   
\usepackage{amsmath, amssymb} 
\usepackage{graphicx}         
\usepackage{algorithm}
\usepackage{algorithmicx}
\usepackage{algpseudocode}
\usepackage{hyperref}         
\usepackage{cite}  
\usepackage{mathrsfs}
\usepackage{geometry}         
\usepackage[numbers]{natbib}
\usepackage{helvet}
\usepackage{subfig}
\usepackage{tabularx}
\usepackage{xr-hyper}
\usepackage{hyperref}
\usepackage{authblk} 

\title{Accelerating Privacy-Preserving Medical Record Linkage: A Three-Party MPC Approach}

\author[1,2]{Şeyma Selcan Mağara\thanks{Corresponding author: seyma-selcan.magara@uni-tuebingen.de}}
\author[1]{Noah Dietrich}
\author[1,2]{Ali Burak Ünal}
\author[1,2]{Mete Akgün}
\date{\vspace{-5ex}}

\affil[1]{Medical Data Privacy and Privacy-Preserving ML on Healthcare Data, Dept. of Computer Science, University of Tübingen}
\affil[2]{Institute for Bioinformatics and Medical Informatics, Tübingen, Germany}
\begin{document}

\maketitle

\begin{abstract}
Record linkage is a crucial concept for integrating data from multiple sources, particularly when datasets lack exact identifiers, and it has diverse applications in real-world data analysis. Privacy-Preserving Record Linkage (PPRL) ensures this integration occurs securely, protecting sensitive information from unauthorized access. This is especially important in sectors such as healthcare, where datasets include private identity information (IDAT) governed by strict privacy laws. However, maintaining both privacy and efficiency in large-scale record linkage poses significant challenges. Consequently, researchers must develop advanced methods to protect data privacy while optimizing processing performance. This paper presents a novel and efficient PPRL method based on a secure 3-party computation (MPC) framework. Our approach allows multiple parties to compute linkage results without exposing their private inputs and significantly improves the speed of linkage process compared to existing privacy-preserving solutions. We demonstrated that our method preserves the linkage quality of the state-of-the-art PPRL method while achieving up to 14 times faster performance. For example, linking a record against a database of 10,000 records takes just 8.74 seconds in a realistic network with 700 Mbps bandwidth and 60 ms latency. Even on a slower internet connection with 100 Mbps bandwidth and 60 ms latency, the linkage completes in 28 seconds, highlighting the scalability and efficiency of our solution.
\end{abstract}


\newcommand{\fw}{CECILIA}
\newcommand{\app}{Sort}

\newcommand{\party}{\mathcal{P}} 
\newcommand{\helper}{Helper} 
\newcommand{\numbit}{\ell} 
\newcommand{\inp}{a} 
\newcommand{\inpvec}{\Vec{a}}
\newcommand{\arrlen}{m} 

\newcommand{\bshare}[4]{\langle {#1^{#2}} \rangle^{{#3}}_{B_{#4}}} 
\newcommand{\ashare}[5]{\langle #1^{#2}_{#5} \rangle^{{#3}}_{A_{#4}}}  
\newcommand{\pshare}[4]{\langle {#1^{#2}} \rangle^{{#3}}_{P_{#4}}}  

\newcommand{\bsharevector}[4]{\langle \Vec{#1}^{#2} \rangle^{{#3}}_{B_{#4}}} 
\newcommand{\asharevector}[5]{\langle \Vec{#1}^{#2}_{#5} \rangle^{{#3}}_{A_{#4}}} 

\newcommand{\vect}[3]{\Vec{#1}^{#2}_{#3}}

\newcommand{\comPerm}{\Vec{\pi}} 
\newcommand{\crn}{r} 
\newcommand{\crb}{\rho} 
\newcommand{\rb}{\Vec{r}} 
\newcommand{\rn}{\Vec{r}}

\newcommand{\permG}{\Vec{\Delta}}
\newcommand{\permC}{\Vec{\delta}}
\newcommand{\perm}[2]{#1 \circ #2} 
\newcommand{\xor}[2]{#1 \oplus #2} 
\newcommand{\sortedbits}{\Vec{\gamma}}

\newcommand{\findallmatch}{\mathrm{FindAllMatches}}
\newcommand{\comprecsim}[2]{\mathrm{CompRecordSimilarities}(#1 , #2)}
\newcommand{\compmaxscore}[1]{\mathrm{CompMaxScores}(#1)}
\newcommand{\getmatch}[1]{\mathrm{GetMatches}(#1)}
\newcommand{\gencomrand}[2]{\mathrm{GenerateCommonRandom}(#1 , #2)}
\newcommand{\multiplication}[2]{\mathrm{Multiply}(#1,#2)}
\newcommand{\compare}[2]{\mathrm{Compare}(#1,#2)}
\newcommand{\multiplex}[3]{\mathrm{Multiplex}(#1,#2,#3)}

\newcommand{\matches}{\mathcal{M}}
\newcommand{\records}{\mathrm{R}}

\newcommand{\simu}{\mathcal{S}} 
\newcommand{\func}[1]{\mathcal{F}_{#1}} 
\newcommand{\adv}{\mathcal{A}}

\newcommand{\threshold}{\lambda}

\newcommand{\maltriple}{k}
\newcommand{\totaltriple}{n}
\newcommand{\mackey}{\kappa}

\newcommand{\ring}{\mathbb{Z}_{2^{64}}}
\newcommand{\ringA}{L} 
\newcommand{\ringB}{K} 
\newcommand{\ringC}{V} 
\newcommand{\ringBool}{B} 
\newcommand{\setR}{\mathbb{E}} 
\newcommand{\setnf}{S} 

\newcommand{\evl}{\Lambda} 
\newcommand{\evc}{Q} 
\newcommand{\clas}{w} 

\newcommand\vartextvisiblespace[1][.5em]{%
  \makebox[#1]{%
    \kern.07em
    \vrule height.3ex
    \hrulefill
    \vrule height.3ex
    \kern.07em
  }
}

\section{Introduction}
Record linkage is the identification of records referring to the same entity across different datasets, even when these records do not share a common identifier. This technique is essential in various fields, especially when merging datasets from different institutions or sources, to get a more comprehensive analysis and gain deeper insights. For example, linking cancer registry data with treatment and follow-up records has been used to assess long-term survival rates and outcomes \cite{Gatta533} or it can be used for identifying the adverse drug reactions for specific patient groups \cite{christen2006privacy}. More recently, linking patient records across hospitals allowed researchers to better understand disease progression and outcomes \cite{2020.05.06.20092999}.

Since its first formulation by \cite{fellegi-sunter}, many advanced techniques have been proposed to link the entities across separate datasets. However, these datasets often contain sensitive personal information, such as patient names or Social Security Numbers (SSNs), which introduces significant challenges in data privacy. Also, legal regulations, such as the European General Data Protection Regulation (GDPR) \cite{gdpr} and the Health Insurance Portability and Accountability Act (HIPAA) \cite{hipaa1996}, prevent the exchange of clear-text personally identifiable information between institutions without the patient's explicit consent.

Several privacy-preserving record linkage (PPRL) methods have been proposed, utilizing encryption and record linkage algorithms to link records without revealing individual identities. One widely used method is Bloom filters \cite{bloom} to enable probabilistic linkage of strings even in the presence of data errors or inconsistencies with the help of a trusted third-party (TTP) \cite{schnell2009, Durham2010PrivateMR, 6529084, RANDALL2014205}. While effective, Bloom filter-based approaches are vulnerable to cryptanalysis and frequency attacks \cite{Christen2017EfficientCO}. To reduce these security vulnerabilities and to maintain low collision rates, Bloom filters require significant space. 

On the other hand, in MPC settings, string similarity can be directly computed on secret-shared bigrams since MPC ensures data privacy through secret shares, eliminating the need for Bloom filters. Hence, Privacy-Preserving Record Linkage (PPRL) can be effectively achieved using secure MPC with bigram-based methods without additional overhead.
While secret-sharing provides robust privacy protections, the state-of-the-art solution, MainSEL \cite{MainSEL} operating within a two-party MPC framework, still uses Bloom filters for data encoding. However, Bloom filter-based methods do not outperform bigram-based methods in terms of efficiency and accuracy, even in MPC settings. This limitation arises from the inherent inefficiency of two-party MPC protocols, which require computationally and communication-intensive operations like Beaver triple generation. In contrast, three-party MPC solutions mitigate these overheads, allowing bigram-based methods to achieve superior performance. Consequently, bigram-based solutions offer a more efficient, precise, and scalable alternative to Bloom filter-based methods for PPRL in secure MPC environments.

This paper proposes a novel three-party record linkage framework that addresses the limitations of Bloom filter-based methods, TTP reliance, and two-party systems. We introduce a bigram mapping approach to compute string similarity directly on secret-shared bigrams, reducing the setup complexity for data owners and the space overhead associated with Bloom filters. Our three-party setting eliminates the need for an offline session, significantly improving the efficiency of the linkage process. Additionally, we achieve faster computation for certain operations than traditional two-party systems by using inherently advantaged three-party MPC systems. The proposed method ensures that no party gains unauthorized access to sensitive data while providing a more secure, efficient, and faster solution compared to state-of-the-art PPRL techniques.

In this work, we present four key contributions that advance the field of PPRL through innovative computational frameworks and methodologies:

\textbf{Three-Party Computation Framework:} 
Our framework improves the runtime performance of the two-party solution MainSEL \cite{MainSEL} using a three-party (aided two-party) computation model. The transition to 3-party setting significantly accelerates the online phase and eliminates the need for a setup phase, typically required in two-party computations. As a result, our protocol is more efficient and less resource-intensive. 

\textbf{Elimination of Bloom Filters:} 
We replace Bloom filters with a direct bigram mapping approach, where string similarity is computed directly on secret-shared bigrams. Unlike Bloom filters used in the MainSEL to encode q-grams, our method securely handles bigrams without additional Bloom filter calculation steps. This enhancement reduces the setup complexity for data owners, requires less space, and eliminates the vulnerabilities associated with Bloom filters, such as frequency attacks and cryptanalysis. As a result, our approach simplifies the process while maintaining privacy in MPC.

\textbf{Better Suitability for Large-Scale Data Linkage:}
Our solution's efficiency improvements make it better suited for large-scale record linkage tasks. Lower computation and communication costs enable scaling to larger datasets, making our framework practical for real-world applications involving substantial data volumes.

\textbf{Support for Multiple Data Owners:}
Our framework inherently supports linkages between any number of data owners. Data owners send their secret shares to the framework’s parties, enabling linkage across multiple databases. This feature expands the applicability of our method to scenarios involving more than two databases, increasing flexibility and usability in practical PPRL applications.

\section{Preliminaries}\label{sec:preliminaries}
\subsection{Multi-Party Computation}\label{sec:multi-party-comp}

Secure Multi-Party Computation (MPC) was pioneered by \cite{Yao} with the introduction of Yao's Garbled Circuits, allowing two parties to compute a function securely without revealing their inputs. Following Yao's work, \cite{gmw} expanded this idea and formulated a comprehensive MPC framework. Later, the SPDZ protocol utilized pre-computed correlated randomness and homomorphic encryption, leading to improved and efficient MPC protocols \cite{damgaard2012multiparty}.

 \cite{Kamara2011OutsourcingMC} introduced an outsourcing model in which a powerful server assists in computations while remaining unaware of the private data. By utilizing an oblivious server, the computational load on participating parties is significantly reduced, enabling efficient and secure processing of sensitive information. This approach has been adopted and further developed in various studies, including works like ABY \cite{demmler2015aby}, SecureML \cite{secureml}, SecureNN \cite{securenn} and other significant frameworks \cite{mohassel2018aby3,wagh2021falcon,cecilia}. Moreover, significant efforts have been made to bring MPC closer to practical, real-world applications, as demonstrated by works such as \cite{malkhi2004fairplay,aby2}. MPC has found significant use in privacy-preserving machine learning and data analysis, demonstrating its importance in maintaining data confidentiality while enabling collaborative computations \cite{secureml, securenn, mohassel2018aby3, patra2020blaze, byali2020flash, wagh2021falcon}.
 
\subsection{Record Linkage}\label{sec:record-linkage}
Record linkage involves matching records from different datasets that belong to the same entity, such as an individual. When a unique identifier, like an SSN, is available, matching the records is theoretically trivial. However, in practice, these unique identifiers are often missing. When they are available, data errors can complicate the matching process, requiring reliance on common attributes like name or birth date found in both datasets, which can also be erroneous. 

In medical research, record linkage involves merging patient data from various sources. These datasets typically contain the patient's name, location and date of birth. However, these identifiers are not unique and are subject to errors, which create significant challenges. These inaccuracies typically arise from human errors during manual data entry, such as typos, spelling mistakes, or misplacing information like switching first and last names.
To address these complexities, three main approaches to record linkage have developed: deterministic, probabilistic, and machine learning-based methods.

Deterministic record linkage relies on strict rules to classify records as matching or not. This method requires high-quality data and is less common due to its inability to handle errors. 

In contrast, probabilistic approach is more flexible, calculating a match probability between record pairs and categorizing them based on a threshold. This method better handles lower-quality data with errors, allowing adjustments between specificity and sensitivity. The most common probabilistic record linkage method is the \cite{fellegi-sunter} (FS) model, which matches records by summing partial match weights for each attribute, adjusted by prior probabilities and error rates. The effectiveness of this approach depends on the specific implementation and calculation of probabilities.

Machine-learning based record linkage has gained importance recently due to their ability to build accurate linkage and to estimate error rates. Early works showed that the FS model is equivalent of naïve Bayes classifier \cite{6033192}. Several works used machine learning algorithms, such as SVMs \cite{10.1145/956750.956759}, neural networks \cite{6033192}, random forest and gradient boosting classifiers \cite{10.1145/2951894.2951896} in pairwise linkage of data. These methods supports dependency between the fields \cite{6033192}, however they may still suffer from machine learning related issues such as missing values \cite{AIKEN2019104857}.

Various probabilistic record linkage solutions have sought to improve the efficiency and usability of the FS model, particularly in settings with resource constraints or where large-scale linkage is required. For example, \cite{contieroepilink} introduced a lightweight probabilistic RL solution designed to operate effectively in environments with limited computational resources. \cite{lablans_restful} implemented this in the Mainzelliste framework, which offers a RESTful web service for record linkage. The framework is currently in use in various medical joint research projects for pseudonymization/depseudonymization of the patient's IDAT. 

\subsection{Related Work}
Privacy-preserving techniques for record linkage are crucial due to the strict data protection regulations and the inherent risks of handling sensitive personal information. Regulations like GDPR in the European Union \cite{gdpr} and HIPAA \cite{hipaa1996} in the United States mandate that organizations protect individuals' privacy when processing personal data. These laws require that data sharing or processing, including record linkage, minimize the risk of re-identification and unauthorized access.

A straightforward approach addressing these requirements is encoding the personal data in a way that enables secure linkage computation. One way of doing this is by calculating the hash of the IDAT and comparing the hash values. One drawback of this method is that hashes are still vulnerable to dictionary attacks. Laud et al. \cite{laud} utilizes MPC and the hashes to compare records. They realized this solution with the Sharemind framework \cite{sharemind} and achieved fast execution time, linking $10000$ records in $30$ minutes in a highly congested network. However, this method relies on an exact matching and requires complete and correct databases. 

While exact matching is efficient in specific scenarios, it has significant limitations when dealing with real-world data. Given the limitations of exact matching, probabilistic approaches offer a more realistic solution in many scenarios. One common technique used for this purpose is Bloom filter, a method proposed by \cite{bloom} initially to check the set membership of an element and they are also used in string similarity calculation. Bloom filters use q-grams to compute the similarity of two fields. First, each field, such as names, is divided into sets of consecutive $q$ letters known as q-grams. If there are no privacy concerns, the similarity between these two fields can be calculated simply by computing the ratio of the common q-grams to total q-grams, known as Dice coefficient. However, privacy-preserving methods require hiding the q-gram values. Schnell et al. \cite{schnell2009} uses bloom filters to meet this requirement by mapping each q-gram into a bit array using multiple cryptographic hash functions. This securely encodes q-grams without revealing the original data. Similarity is then assessed by computing the Dice coefficient on the bit arrays. 
Since this work, Bloom filters have been widely used in PPRL methods to compute probabilistic matching in string fields.

In \cite{schnell2009}, a TTP receives the bloom filters and computes the dice coefficients, assuming that no sensible information is deducible from bloom filters. However, it has shown that Bloom filters are vulnerable to frequency attacks \cite{Christen2017EfficientCO}. One proposed method to overcome this attack is combining the MPC into bloom filters. The work of \cite{lazrig2018}, based on \cite{schnell2009}, mitigates this risk by using Yao's Garbled Circuits and eliminating the TTP.

Building upon the progress made in PPRL, the MainSEL \cite{MainSEL} method represents the current state-of-the-art. It was developed as an MPC variant of the Mainzelliste \cite{lablans_restful} using the ABY framework \cite{demmler2015aby}. MainSEL addresses the information leakage problem present in Lazrig's method \cite{lazrig2018}, where Bloom filter matches could potentially reveal sensitive information. Additionally, \cite{MainSEL} introduced a novel technique for resolving ties between multiple probabilistic matches by computing the highest match score within MPC, a capability absent in Lazrig et al.'s solution \cite{lazrig2018}. Furthermore, MainSEL compares fields like date of birth or zip code using exact equality and combines these comparisons into a weighted sum to determine the final score, while Lazrig et al.'s solution \cite{lazrig2018} only performs threshold comparisons on individual fields.

\section{Method}\label{sec:method}
This section describes our record linkage protocol, starting with the traditional non-private approach and then transitioning to the proposed privacy-preserving method based on MPC. 

\subsection{Non-private Record Linkage Protocol} \label{sec:plain-record-linkage}
To compute the matches between two databases $X=\{x^1, x^2, ... x^n\}$ and $Y=\{y^1, y^2, ... y^n\}$, first, a similarity score should be calculated between each pair ($x^i$, $y^j$), where $i$, $j$ are the record indices. A field in records is denoted by $X_f$, $f\in F$ and $F$ is the field index. A record includes both fuzzy fields (such as name, surname, birth name, and city) and exact fields (such as postcode, birth year, month, and day). The similarity of each field is calculated separately. For exact fields, a direct comparison is performed, resulting in 1 if the fields are equal or 0 otherwise. Thus, for two records $x^i$ and $y^j$, the similarity of an exact field $f$ is calculated as
$sim(x_f^i, y^j_f) = (x^i_f == y^j_f)$

\begin{figure}[]
    \centering
    \includegraphics[width=0.85\linewidth]{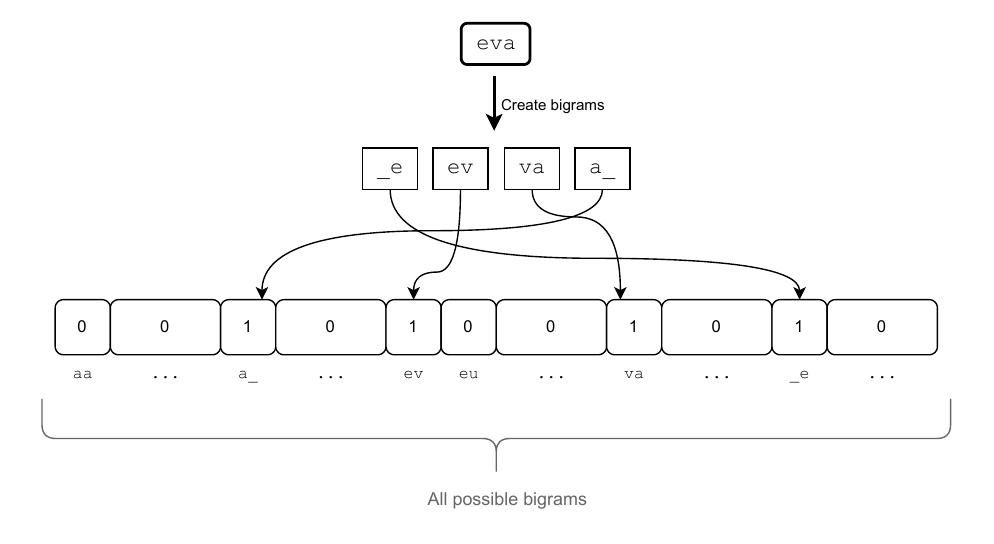}
    \caption{Mappings of bigrams}
    \label{fig:bin-encod-string}
\end{figure}

For fuzzy fields, such as the string fields, a similarity check is preferable to an equality check as it accommodates data with errors. Similarity check is using bi-grams of the string and the Dice coefficient score. First the bi-grams in a fuzzy field $x^i_f$ are identified andan array indicating the presence of bi-grams is created, as illustrated in Figure \ref{fig:bin-encod-string}. The resulting array, referred as bi-gram mapping and denoted as $M(x^i_f)$, has a fixed length corresponding to the number of all possible bi-grams. Given an alphabet $\Sigma$, where $\Sigma = \{a,b,c,...,z,-,.,\vartextvisiblespace,*\}$, $M(x^i_f)$ is an array of size $900$. The elements of $M(x_f)$ are initially set to 0. Then, for each bi-gram found in the string, the corresponding index of $M(x^i_f)$ is set to one. Once the bi-gram mappings of the same fuzzy field $f$ of two records $x^i$ and $y^j$ are computed, their similarity based on this fuzzy field is calculated using the Dice coefficient, as shown in Equation \ref{eq:dice}.

\begin{equation} \label{eq:dice}
        \text{sim}(x^i_f, y^j_f) = \frac{2*|M(x^i_f) \cap M(y^j_f)|}{|M(x^i_f)| + |M(y^j_f)|}
\end{equation}

The numerator, $|M(x^i_f) \cap M(y^j_f)|$, counts the number of common bi-grams between the two mappings. This value is multiplied by 2 to balance the size of both sets in the formula. The denominator, $|M(x^i_f)| + |M(y^j_f)|$, is the sum of the total bi-grams in each field.

After calculating the similarities of individual fields, they are used to compute the overall similarity score of the record pair. Each field has a predetermined weight, calculated based on the frequency and the error rates as suggested in \cite{contieroepilink}. However, as they mentioned, these values can be found by trial-and-error or derived from previous databases. 

The normalised weighted sum of the field similarities of a pair gives the final similarity score as shown in Equation \ref{eq:sim-score}. In the equation, $w_f$ is the weight of the field $f$, and $\delta_f$ is the value denoting whether the field is non-empty (1) or not (0). However, for the name field, which combines first, last, and birth names, $\delta_f$ is adjusted based on the number of missing components, such as $\delta=0.66$ if one is missing. This ensures that the weight is adjusted proportionally to the completeness of the information in the name field.

\begin{equation}\label{eq:sim-score}
    S(x^i, y^j)=\frac{\sum_{f \in F} \delta_{f} w_f \operatorname{sim}\left(x^i_f, y^j_f\right)}{\sum_{f \in F}\delta_{f} w_f}
\end{equation}


Once the similarity scores between a record $x^i$ and all records in the database $Y$ are computed, the next step is to identify the record with the highest score. Using the argmax function, the record $y^{\epsilon}$ that maximizes the similarity score is selected, as shown in Equation \ref{eq:max-match}. If the similarity score of $y^{\epsilon}$ exceeds the predefined threshold $\tau$, $y^{\epsilon}$ is considered a match. Otherwise, no match is found.

\begin{equation}\label{eq:max-match}
\begin{split}
    \epsilon = argmax_{i \in N} S(x^k,y^i) \\
        match = 
\begin{cases}
    y^{\epsilon} & \text{if } S(x^k, y^{\epsilon}) > \tau \\
    null,              & \text{otherwise}
\end{cases}
\end{split}
\end{equation}

\subsection{Privacy-preserving Record Linkage Protocol}\label{sec:ppml}
This section explains the framework and its utilization in our implementation and the secret sharing process, as well as the privacy-preserving record linkage protocol.

\subsubsection{Framework}\label{sec:fw}
The proposed approach for record linkage is implemented using a secure MPC framework which utilizes a 2-out-of-2 additive secret-sharing in $\mathbb{Z}_{2^{64}}$. In this system, a value is split into two additive shares such that neither share reveals any information about the original value unless combined with the other. For example, if a value $a$ is to be secret shared, a random $r\in\mathbb{Z}_{2^{64}}$ is generated, and the shares are defined as $\langle a \rangle_0 = r \mod \mathbb{Z}_{2^{64}}$  and $\langle a \rangle_1 = (a-r) \mod \mathbb{Z}_{2^{64}}$. Each party holds one of these shares, ensuring the original value remains private.
\begin{figure}[htp]
    \centering
    \includegraphics[width=0.8\linewidth]{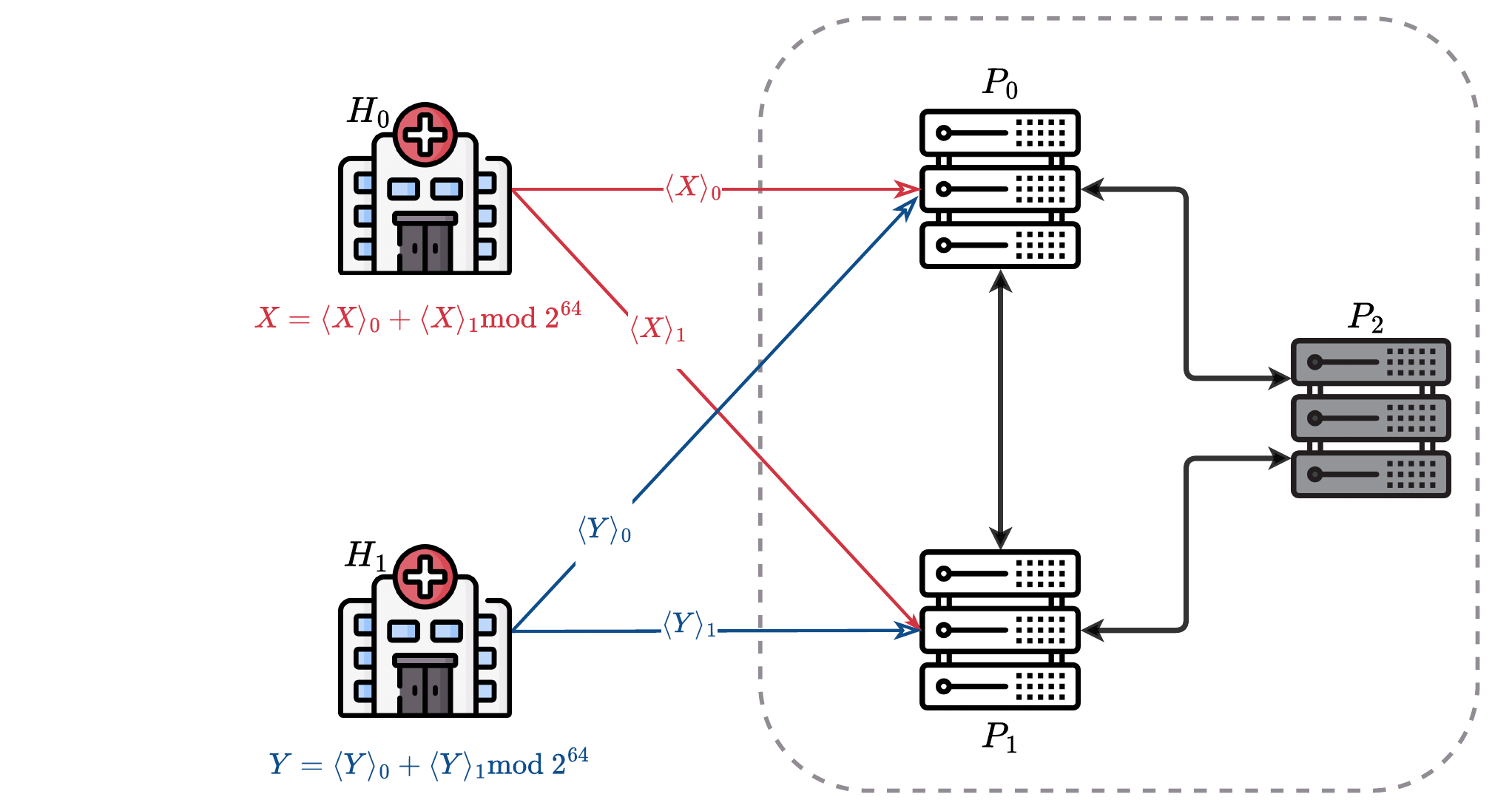}
    \caption{Framework and secret sharing of the records}
    \label{fig:cecilia}
\end{figure} 
As depicted in Figure \ref{fig:cecilia}, the framework involves three parties: $\party_0$ and $\party_1$ (referred to as proxies) and $\party_2$ (referred to as the $\helper$ party); and the proxies hold additive secret shares of the records to be linked. The $\helper$ party provides the proxies with shares of specific values, such as beaver triples, and assists the proxies in computing a desired function. Since the $\helper$ performs computations on masked data and returns shares of the result, privacy is not compromised.

\subsubsection{Security Model}\label{sec:sec-model}
We follow a semi-honest security model with an honest majority, meaning all parties follow the protocol as prescribed but may try to learn information from the data they receive. However, using MPC ensures that inputs from all parties are kept private by arithmetically sharing the private input. None of the individual shares sent to parties alone provide meaningful information about the original input. The servers perform computations on these shares throughout the process without gaining knowledge of the underlying data.

Proxies conceal the values of their shares by adding randomness when sending them to the $\helper$ within the protocol. This randomness ensures that the information received appears random to the $\helper$, preventing the $\helper$ extracting any information from the input data.

Information such as database size or number of fields is given to proxies without secret sharing. However, if they need to be hidden, they can be easily hidden with padding methods.

In summary, the framework guarantees that no private information is leaked during the record linkage process, preserving the confidentiality of sensitive data even in the presence of up to one corrupted server. The final output, whether a match is found, is revealed only to the parties entitled to receive this result, while all other information remains protected.

\subsubsection{Secret Sharing of the Records: Outsourcing}\label{sec:record-sharing}
In practice, our record linkage system can involve multiple dataholders, each contributing their own dataset. To illustrate the protocol more clearly, we simplify by assuming $H_0$ and $H_1$ are two data holders, with the records of the first one being matched to the second database. However, this process can be extended to involve multiple parties, as proxies can interact with the data owners and obtain their data shares to privately find the matches.

$H_0$ and $H_1$ hold records $X$ and $Y$ respectively. A record includes both fuzzy fields (name, surname, birth name, and city) and exact fields (postcode, birth year, month, and day). Each data owner generates shares of their records. Shares of the exact fields are created directly. For the fuzzy fields, the data owners first identify the bigrams and create an array indicating the bigrams, as explained in \nameref{sec:plain-record-linkage} section. 

The first name, birth name, and surname are combined and processed together: Data owners concatenate these fields and calculate the bi-grams over the combined name. This approach helps to handle a common entry error where the name and surname might be mistakenly swapped in the database.  A separate encoding vector is created for the city field. 

Data owners also calculate the number of bi-grams in the combined name and city fields. Then, the data owners create the shares of these arrays and the number of bi-grams and the shares of the exact fields. They send one share of their records to $\party_0$ and the other to $\party_1$ as shown in Figure \ref{fig:cecilia}. As a result, each party gets one share of each record. After this, proxies proceed to privacy-preserving record linkage protocol.

\subsubsection{Privacy-Preserving Record Linkage}\label{sec:sec-record-linkage}
Our method employs a three-party MPC framework to perform record linkage securely, eliminating the need for Bloom filters and enhancing both privacy and efficiency. The approach leverages a bigram mapping technique combined with secure computation protocols to calculate string similarity directly, significantly reducing communication costs and improving execution speed.

We begin by outlining the overall process of our record linkage algorithm, as depicted in \textbf{Algorithm \ref{alg:recordlinkage}}. Upon receiving the shared records, the proxies execute the following steps:

\begin{algorithm}
\scriptsize
\caption{FindAllMatches}\label{alg:recordlinkage}
\begin{algorithmic}[1]
\Require Record shares $\langle x_i \rangle$ and $\langle Y \rangle$ 
\Ensure $\langle \matches \rangle$: sharing of matching records of high score
\State $\langle S_r \rangle \Leftarrow \mathrm{CompRecordSimilarities}(\langle x^i \rangle, \langle Y \rangle)$
\State $\langle S_m \rangle \Leftarrow \mathrm{CompMaxScore}(\langle S_r \rangle)$
\State $\langle \matches \rangle \Leftarrow \mathrm{GetMatches}(\langle S_m \rangle)$
\end{algorithmic}
\end{algorithm}

First, the `CompRecordSimilarities` function calculates pairwise similarity scores between record $x^i$ and database $Y$. This involves computing both exact and fuzzy field similarities. The fuzzy field similarity is determined using the Dice coefficient, as implemented in \textbf{Algorithm \ref{alg:diceSim}}, while exact fields utilize a straightforward equality check demonstrated in \textbf{Algorithm \ref{alg:exactSim}}.

\begin{algorithm}
\scriptsize
\caption{DiceSimilarity}\label{alg:diceSim}
\begin{algorithmic}[1]
\Require Bigram map $\langle M(x^i_f) \rangle$, $\langle M(y^j_f) \rangle$ 
\Require Bigram cardinality shares $\langle |M(x^i_f)| \rangle$, $\langle |M(y^j_f)| \rangle$
\Ensure Dice score similarity $\text{sim}(x^i_f,y^j_f)$ of the records
\State $\langle b \rangle \Leftarrow \mathrm{And}(\langle M(x^i_f) \rangle, \langle M(y^j_f) \rangle)$
\State $\langle n \rangle \Leftarrow 2 \cdot \sum_{l=1}^{900} \langle b \rangle[l]$
\State $\langle d \rangle \Leftarrow \langle |M(x^i_f)| \rangle + \langle |M(y^j_f)| \rangle$
\State $\langle S_f \rangle \Leftarrow (\langle n \rangle, \langle d \rangle)$
\State \textbf{return} $S_f$
\end{algorithmic}
\end{algorithm}

In \textbf{Algorithm \ref{alg:diceSim}}, the Dice coefficient for fuzzy fields is securely computed by performing a bitwise AND on the shared bigram maps to identify common bigrams. The intersection size is then summed and doubled to form the numerator, while the sum of the cardinalities of both bigram maps constitutes the denominator of Equation \ref{eq:dice}. These secret-shared values are returned as separate entities to avoid costly division operation within the MPC framework.

\begin{algorithm}
\scriptsize
\caption{ExactSimilarity}\label{alg:exactSim}
\begin{algorithmic}[1]
\Require Shares of exact fields $\langle x^i_f \rangle$, $\langle y^j_f \rangle$ 
\Ensure Exact similarity $\text{sim}(x^i_f,y^j_f)$ of the records
\State $\langle S_e \rangle \Leftarrow \mathrm{Equals}(\langle x^i_f \rangle, \langle y^j_f \rangle)$
\State \textbf{return} $\langle S_e \rangle$
\end{algorithmic}
\end{algorithm}

For exact fields such as zip codes, \textbf{Algorithm \ref{alg:exactSim}} performs an equality check by comparing the shared values and assigning a similarity score of 1 for matches and 0 otherwise.

After computing the similarities, the `CompRecordSimilarities` function aggregates these scores using a normalised weighted average, as shown in \textbf{Algorithm \ref{alg:recordSim}}. This weighted sum combines both fuzzy and exact field similarities, normalizing by the total weight of the fields. Since the weights are not secret, multiplication and division by the weights are performed locally. The weighted sum operates on the pre-computed numerators and denominators carried forward from the fuzzy field similarity calculation. As a result, our method returns the numerator and denominator of the final weighted similarity score, allowing us to efficiently perform the normalization without a costly division operation.

\begin{algorithm}[H]
\scriptsize
\caption{CompRecordSimilarities}\label{alg:recordSim}
\begin{algorithmic}[1]
\Require Exact field shares $\langle x^i_f \rangle$, $\langle y^j_f \rangle$ 
\Require Bigram map $\langle M(x^i_f) \rangle$, $\langle M(y^j_f) \rangle$ 
\Require Bigram cardinality shares $\langle |M(x^i_f)| \rangle$, $\langle |M(y^j_f)| \rangle$
\Ensure Similarity score $\langle S_r \rangle$ of the records
\State $\langle S_f \rangle \Leftarrow \mathrm{DiceSimilarity}(\langle M(x^i_f) \rangle, \langle M(y^j_f) \rangle)$ for each fuzzy field $f$
\State $\langle S_e \rangle \Leftarrow \mathrm{ExactSimilarity}(\langle x^i_f \rangle, \langle y^j_f \rangle)$
\State $\langle S_r \rangle \Leftarrow \mathrm{WeightedAverage}(\langle S_f \rangle, \langle S_e \rangle, W)$
\State \textbf{return} $\langle S_r \rangle$
\end{algorithmic}
\end{algorithm}

Once similarity scores are computed, the next step involves identifying the highest scoring matches using \textbf{Algorithm \ref{alg:computemax}}. This algorithm compares the scores through cross-multiplication of their numerator and denominator shares to avoid expensive division operations. The `Compare` and `Multiplex` functions within the MPC framework are utilized to securely determine and select the higher score.

\begin{algorithm}[H]
\scriptsize
\caption{ComputeMaxScores}\label{alg:computemax}
\begin{algorithmic}[1]
\Require $\langle S_1 \rangle = (\langle n_1 \rangle, \langle d_1 \rangle)$ and $\langle S_2 \rangle = (\langle n_2 \rangle, \langle d_2 \rangle)$
\Ensure $\langle S_m \rangle = \text{Max}(\langle S_1 \rangle, \langle S_2 \rangle)$
\State $\langle n_1d_2 \rangle \Leftarrow \mathrm{Multiply}(\langle n_1 \rangle, \langle d_2 \rangle)$
\State $\langle n_2d_1 \rangle \Leftarrow \mathrm{Multiply}(\langle n_2 \rangle, \langle d_1 \rangle)$
\State $\langle \gamma \rangle \Leftarrow \mathrm{Compare}(\langle n_1d_2 \rangle, \langle n_2d_1 \rangle)$
\State $\langle S_m \rangle \Leftarrow \mathrm{Multiplex}(\langle n_1d_2 \rangle, \langle n_2d_1 \rangle, \langle \gamma \rangle)$
\end{algorithmic}
\end{algorithm}

Finally, \textbf{Algorithm \ref{alg:getMax}} ensures that only those record pairs with similarity scores exceeding a predefined threshold are considered valid matches. This verification step prevents false positives by comparing the highest score against the threshold using secure comparison functions.

\begin{algorithm}[H]
\scriptsize
\caption{GetMax}\label{alg:getMax}
\begin{algorithmic}[1]
\Require $\langle S_m \rangle$ and threshold $\tau$
\Ensure Matching record $\langle \matches \rangle$ if it exists
\State $\langle \epsilon \rangle \Leftarrow \mathrm{Compare}(\langle S_m \rangle, \tau)$
\State $\langle \matches \rangle \Leftarrow \mathrm{Multiplex}(\langle S_m \rangle, \langle 0 \rangle, \langle \epsilon \rangle)$
\State \textbf{return} $\langle \matches \rangle$
\end{algorithmic}
\end{algorithm}

In \textbf{Algorithm \ref{alg:getMax}}, the maximum similarity score is compared against a threshold using the `Compare` function. If the score exceeds the threshold, the corresponding records are marked as a match using the `Multiplex` function.

\subsection{Code Availability}

The underlying code and datasets for this study is available in PPRL repository and can be accessed via this link \hyperlink{https://github.com/mdppml/PPRL.git}{https://github.com/mdppml/PPRL.git}.

\section{Results}\label{sec:results}
We evaluated the proposed PPRL method in terms of timing and linkage quality, comparing it to the state of the art. We report the results for the linkage quality in terms of false positive and false negative linkages, highlighting how our method performs relative to the existing solution. Additionally, we conducted a comprehensive analysis of execution times to assess the efficiency of our solution compared to a SoTa method.

We ran the experiments on Google Cloud c2-standard-60 machines over Ubuntu 20.04 LTS. Each machine has 60 vCPUs, 240 GB memory. The servers were located in three distinct geographical regions: us-east1-b (East Coast, USA), us-west1-b (West Coast, USA), and us-central1-a (Central USA). The network conditions between the servers were as follows: the bandwidth (BW) between east and central, as well as between west and central, was approximately 700 Mbps, while between east and west, it was around 330 Mbps. Latencies measured were 63 ms between east and west, and 32 ms between central and both other regions.

The default setup simulates a distributed system across different regions, reflecting realistic latency and bandwidth constraints between the parties. Additionally, we limited the bandwidth between parties to 100 Mbps to simulate a slower network to observe the performance under constrained bandwidth conditions typical of slower internet connections. Finally, all computations were run on machines within the same geographical location, with no bandwidth or latency restrictions to simulate a local network environment. The summary of the network settings we used are as follows:

\begin{itemize}
    \item \textbf{Network A}: 32 ms latency and 700 Mbps bandwidth between central and other regions; 63 ms latency and 330 Mbps bandwidth between east and west.
    \item \textbf{Network B}: 32 ms latency between central and other regions; 63 ms latency between east and west; 100 Mbps bandwidth between all regions (slow network simulation).
    \item \textbf{Network C}: 0.1 ms latency and 25 Gbps bandwidth between all parties (local network simulation).
\end{itemize}

\subsection{Dataset} \label{sec:dataset}
Synthetic datasets were created using GeCo \cite{geco} to test the record linkage algorithm. Each record includes first and last names, birth name, birth date (year, month, and day), city, and postcode. First, last, and birth names were generated using GeCo's default frequency files. Frequency tables for cities and postal codes were derived from grid-based analyses of Germany's 2011 census \cite{zensus2011}. Birth dates were generated using a uniform distribution across all days in a year, with birth years based on frequency data from the Statistisches Bundesamt's population pyramid \cite{destatis_pyramide}. 

We applied data corruption using various techniques, including random edits, misspellings, keyboard typos (replacing characters with adjacent keys), and phonetic swaps (swapping similar-sounding character combinations). Two corrupted duplicates were created for each original record, forming two sets of duplicates. Corruption was applied randomly to each attribute with a 10\% overall probability, allowing up to two errors per record. Custom corruption types, such as missing attributes, were also added. Birth names were randomly omitted in 60\% of the records. 

After creating duplicates, 10\% of the records had one of their attribute groups shuffled. Finally, two sets of 10,000 records each were sampled from the duplicates, ensuring that no exact duplicates existed across both sets. The overlap between the sets was set at 60\%, meaning 6,000 records were shared between the two sets.

The weights of the fields are calculated using the the frequency and the error rate of the fields following \cite{contieroepilink}. The weights used can be found in the Supplementary.

\subsection{Linkage Quality}\label{sec
} To evaluate the quality of the record linkage method, the two datasets described in the \nameref{sec:dataset} section were linked without applying any thresholds in the final step, resulting in the highest score and corresponding index for each record in the first database. The same datasets were also used to evaluate the state-of-the-art method \cite{MainSEL}, with updated weights. Since different threshold values serve different needs, it is suggested to explore multiple thresholds \cite{contieroepilink}. Therefore, we analyzed the false positive (FP), false negative (FN), and total error rates across various threshold values.

\begin{table}[!ht]
\centering
\caption{False-Positive (FP) and False-Negatives (FN) in different threshold ($\tau$) values.}
\label{tab:acc-comp}
\begin{tabularx}{\linewidth}{p{0.14\linewidth}|>{\centering\arraybackslash}X|>{\centering\arraybackslash}X|>{\centering\arraybackslash}X|>{\centering\arraybackslash}X|>{\centering\arraybackslash}X|>{\centering\arraybackslash}X|}
\cline{2-7}
 &
  \multicolumn{3}{c|}{MainSEL} &
  \multicolumn{3}{c|}{Ours} \\ \cline{1-7} 
\multicolumn{1}{{|p{0.14\linewidth}|}}{\centering Threshold $\tau$} &
FP &
FN &
Total Errors &
FP &
FN &
Total Errors \\ \hline
\multicolumn{1}{|c|}{$0.60$} &
  \multicolumn{1}{c|}{1146} &
  \multicolumn{1}{c|}{4} &
  \multicolumn{1}{c|}{1200} &
  \multicolumn{1}{c|}{764} &
  \multicolumn{1}{c|}{1} &
  \multicolumn{1}{c|}{\textbf{765}} \\ \hline
\multicolumn{1}{|c|}{$0.65$} &
  \multicolumn{1}{c|}{497} &
  \multicolumn{1}{c|}{11} &
  \multicolumn{1}{c|}{508} &
  \multicolumn{1}{c|}{351} &
  \multicolumn{1}{c|}{4} &
  \multicolumn{1}{c|}{\textbf{355}} \\ \hline
\multicolumn{1}{|c|}{$0.70$} &
  \multicolumn{1}{c|}{54} &
  \multicolumn{1}{c|}{26} &
  \multicolumn{1}{c|}{\textbf{80}} &
  \multicolumn{1}{c|}{148} &
  \multicolumn{1}{c|}{12} &
  \multicolumn{1}{c|}{160} \\ \hline
\multicolumn{1}{|c|}{$0.75$} &
  \multicolumn{1}{c|}{18} &
  \multicolumn{1}{c|}{43} &
  \multicolumn{1}{c|}{\textbf{61}} &
  \multicolumn{1}{c|}{45} &
  \multicolumn{1}{c|}{23} &
  \multicolumn{1}{c|}{68} \\ \hline
\multicolumn{1}{|c|}{$0.80$} &
  \multicolumn{1}{c|}{6} &
  \multicolumn{1}{c|}{339} &
  \multicolumn{1}{c|}{345} &
  \multicolumn{1}{c|}{11} &
  \multicolumn{1}{c|}{51} &
  \multicolumn{1}{c|}{\textbf{62}} \\ \hline
\multicolumn{1}{|c|}{$0.85$} &
  \multicolumn{1}{c|}{2} &
  \multicolumn{1}{c|}{648} &
  \multicolumn{1}{c|}{650} &
  \multicolumn{1}{c|}{1} &
  \multicolumn{1}{c|}{168} &
  \multicolumn{1}{c|}{\textbf{169}} \\ \hline
\multicolumn{1}{|c|}{Best$^{*}$} &
  \multicolumn{1}{c|}{18} &
  \multicolumn{1}{c|}{39} &
  \multicolumn{1}{c|}{57} &
  \multicolumn{1}{c|}{19} &
  \multicolumn{1}{c|}{37} &
  \multicolumn{1}{c|}{\textbf{56}} \\ \hline
  \multicolumn{7}{l}{\scriptsize $^{*}$ The best thresholds are 0.7468 for MainSEL and 0.7852 for our solution.}
\end{tabularx}%
\end{table}

Table \ref{tab:acc-comp} shows that at lower thresholds, both methods exhibit higher false positives and lower false negatives. For instance, at a threshold of 0.60, our method has 764 false positives and 1 false negatives, resulting in a total of 765 errors. In contrast, MainSEL has a total of 1200 errors at the same threshold, with 1146 false positives and 4 false negatives. See Supplementary for more details on FN-FP values.

As the threshold increases, false positives decrease while false negatives increase for both methods, reflecting the stricter matching criteria. Notably, at a threshold of 0.80, our method outperforms MainSEL by achieving a lower total error count of 62 compared to MainSEL's 345. This suggests that our method maintains better overall performance at higher thresholds, effectively reducing incorrect linkages without substantially increasing missed matches.

The optimal thresholds, yielding the minimal total errors, are 0.7468 for MainSEL and 0.7852 for our method. At these thresholds, our method achieves the lowest total errors 56 compared to MainSEL's 57, demonstrating a slight improvement in overall linkage accuracy. This indicates that our approach effectively balances the trade-off between precision and recall across different thresholds.

In addition to analyzing FP, FN, and total errors, we evaluated the performance of both methods using the ROC AUC (Receiver Operating Characteristic Area Under the Curve) metric. MainSEL achieved an ROC AUC of $0.9992$, while our method achieved $0.9997$, indicating exceptionally high accuracy for both methods.

These results demonstrate that our method maintains, and in some cases improves upon, the high linkage quality of the state-of-the-art method. Importantly, these accuracy improvements are achieved without compromising efficiency, making our method suitable for time-sensitive applications requiring both speed and high performance.

\subsection{Runtime Analysis}\label{sec:timings}
\begin{figure*}[ht]%
    \centering
    \subfloat{{\includegraphics[width=0.50\linewidth]{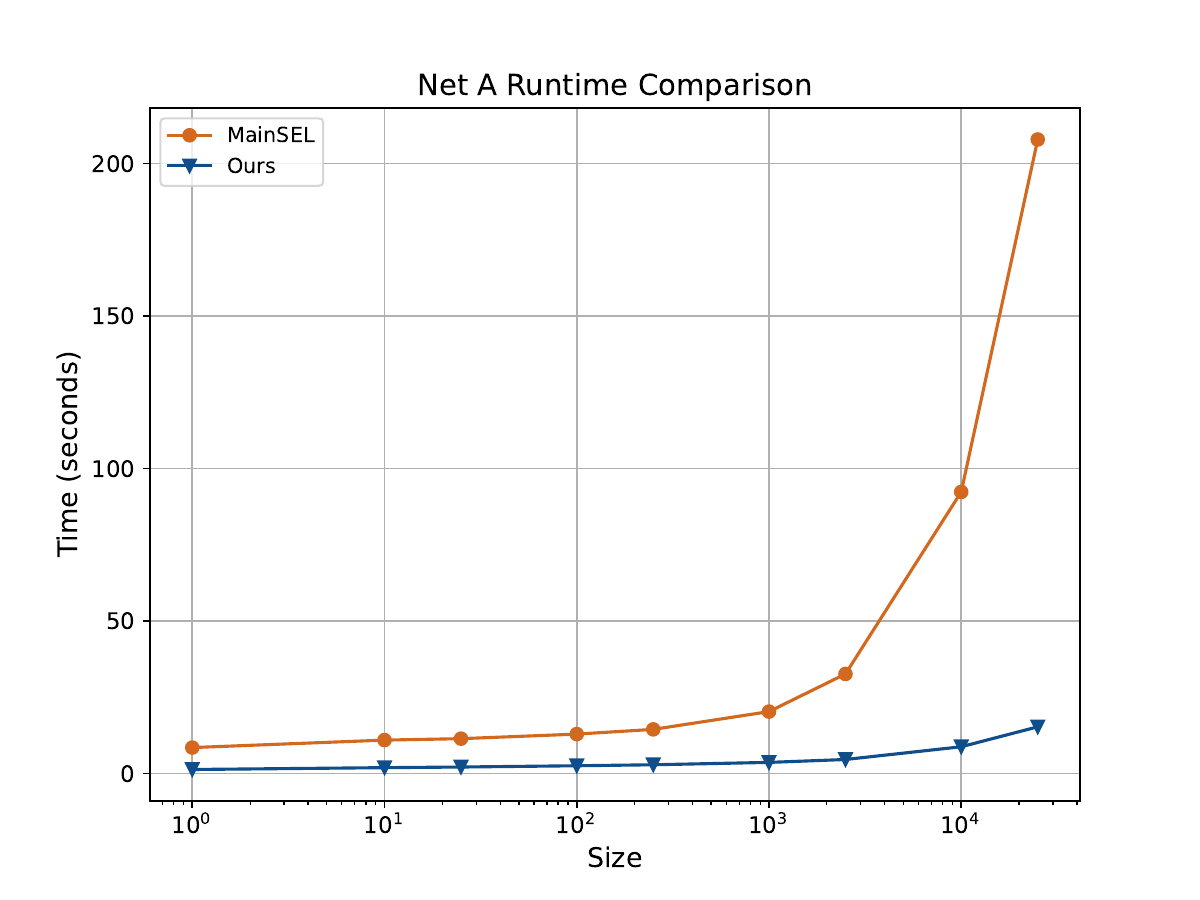} }{\includegraphics[width=0.50\linewidth]{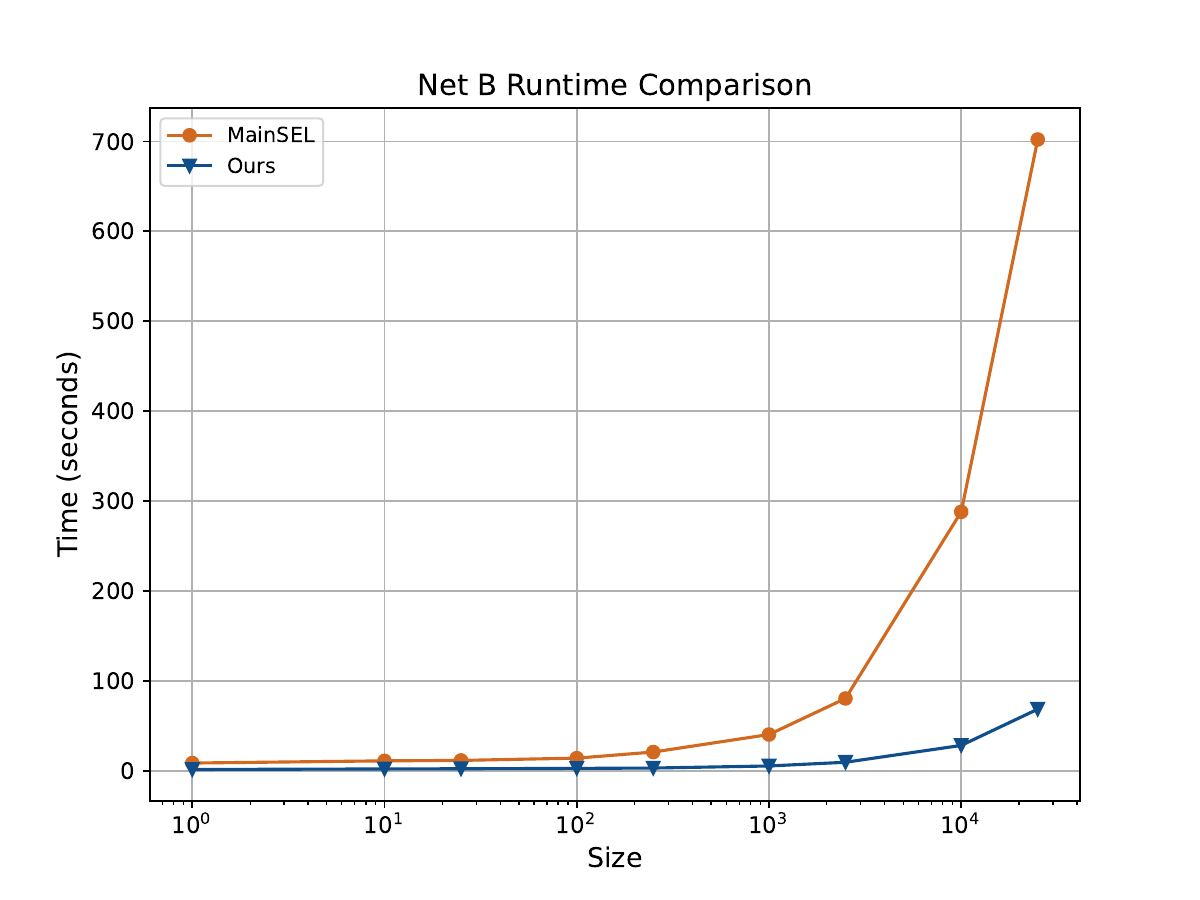} }}\\%
    \subfloat{{\includegraphics[width=0.50\linewidth]{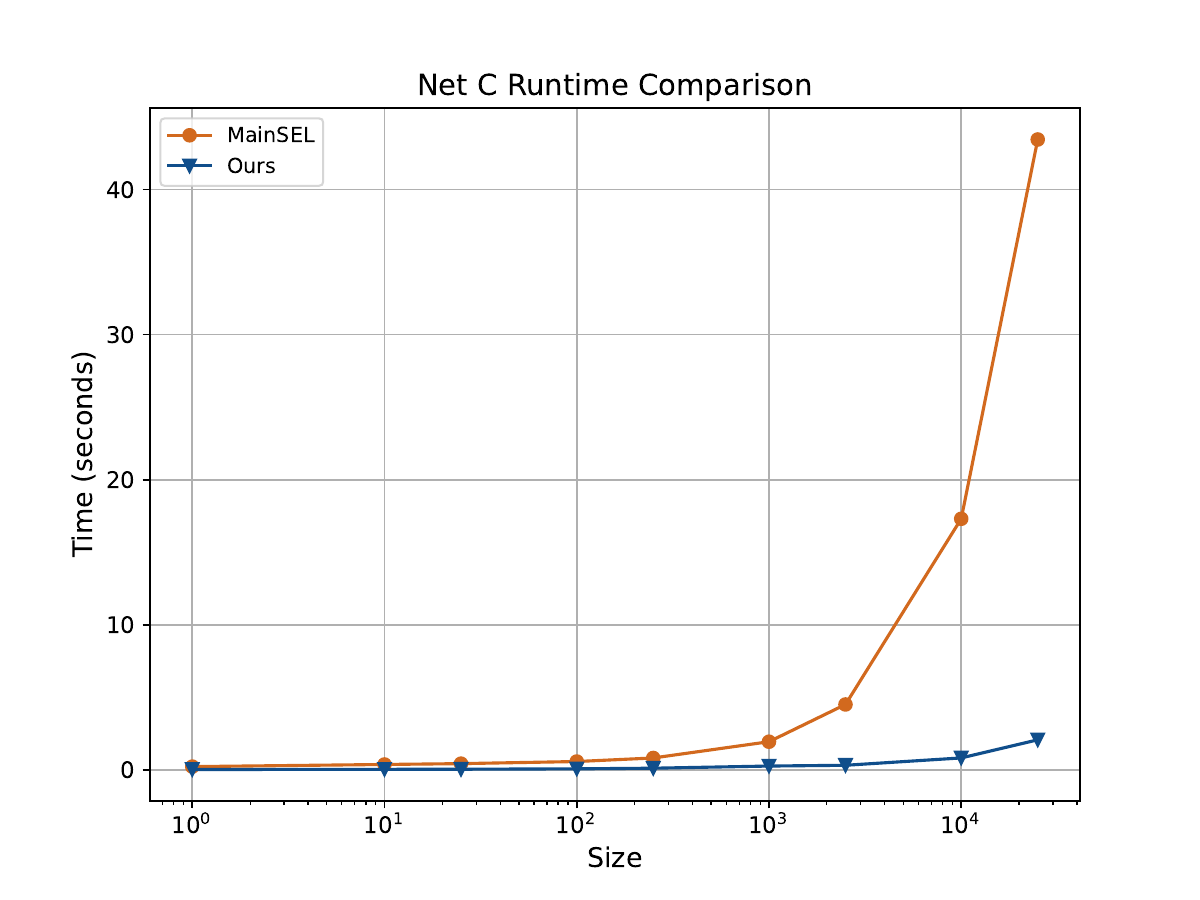} }}%
    \caption{Runtime comparison in different network settings and varying database sizes}%
    \label{fig:timing}%
\end{figure*}

We tested the runtimes of a single record linkage across varying database sizes (from 1 to 25,000 records) under three different network settings. For comparison, we executed the GMW/A circuit variant of MainSEL, demonstrated to give the best performance in \cite{MainSEL}, using the same datasets and within the same testing environment to assess the effectiveness of our solution against state-of-the-art methods. Figure \ref{fig:timing} visualizes the runtime changes as the database size increases for both methods, and the data is also provided in Table \ref{tab:time-comp} for detailed comparison.

The results in Table \ref{tab:time-comp} demonstrate significant improvements in our proposed solution's runtime and communication cost compared to MainSEL across various database sizes and network settings. In all three network conditions (NET A, NET B, NET C), our method consistently outperforms the existing solution, achieving speedups of up to 12x for larger datasets. For instance, in NET A, linking a single record against a database of 10,000 records takes 8.74 seconds with our method, compared to 92.32 seconds using MainSEL. This improvement is even more pronounced in NET B, where our method processes the same operation in 28.09 seconds, compared to 287.96 seconds for MainSEL. The communication cost is also drastically reduced, particularly for larger datasets. For example, linking against 10,000 records in our approach consumes 335.37 MB of communication, contrasting the 6,036.8 MB required by MainSEL.

\begin{table}[!htp]
\centering
\caption{Runtime (in seconds) and communication cost (in MB) comparison of a single record linkage to various size databases in 3 different network settings.}
\label{tab:time-comp}
\begin{tabular}{p{.055\columnwidth}|p{.07\columnwidth}p{.075\columnwidth}|p{.075\columnwidth}p{.075\columnwidth}|p{.06\columnwidth}p{.06\columnwidth}|p{.105\columnwidth}p{.09\columnwidth}}
      & \multicolumn{2}{c|}{NET A} & \multicolumn{2}{c|}{NET B} & \multicolumn{2}{c}{NET C} & \multicolumn{2}{c}{Comm. Cost}\\ \hline
Size & \cite{MainSEL}       & Ours      & \cite{MainSEL}       & Ours      & \cite{MainSEL}       & Ours & \cite{MainSEL}       & Ours     \\ \hline
1        & 8.47        & 1.28 & 8.48   & 1.30 & 0.22 & 0.03 & 0.7 & 0.04\\
10       & 10.92       & 1.88 & 10.93  & 1.91 & 0.37 & 0.04 & 6.2 & 0.34\\
25       & 11.39       & 2.09 & 11.41  & 2.12 & 0.43 & 0.05 & 15.2 & 0.84\\
100      & 12.89       & 2.51 & 13.94  & 2.53 & 0.57 & 0.07 & 60.4 & 3.36\\
250      & 14.44       & 2.83 & 20.71  & 2.92 & 0.82 & 0.11 & 150.7 & 8.39\\
1K     & 20.27       & 3.61 & 40.20  & 5.22 & 1.94 & 0.26 & 602.3 & 33.54\\
2.5K     & 32.62       & 4.56 & 80.31  & 9.36 & 4.51 & 0.32 & 1513.6 &83.85 \\
10K    & 92.32       & 8.74 & 287.96 & 28.09 & 17.31 & 0.82 & 6036.8 & 335.37\\ 
25K    & 207.91      & 15.27 & 702.09 & 68.39 & 43.45 & 2.07 & 15068.2 & 838.43\\ 
\hline\end{tabular}
\end{table}

Additionally, we compared the runtime for complete database linkage for a more realistic evaluation. In a 10,000-by-10,000 record matching, our method completed the task in approximately 24.2 hours in NET A, whereas \cite{MainSEL} required 256.44 hours for the same operation. This drastic reduction emphasizes the superior efficiency of our method, especially in large-scale scenarios.

\section{Conclusion}

We introduced an efficient PPRL method that addresses the performance limitations of state-of-the-art solution while maintaining high accuracy in matching records across datasets. Utilizing a three-party framework, our method eliminates the need for an offline session—a significant bottleneck in state-of-the-art solution—and leverages the uses efficiencies of three-party MPC systems. By replacing computationally intensive Bloom filter operations with a simple q-gram mapping technique, we significantly improve performance without compromising data privacy or linkage quality. Our experiments show that our solution maintains high linkage accuracy and slightly outperforms the state-of-the-art method in ROC-AUC value, demonstrating that our speed enhancements do not sacrifice accuracy. Moreover, our runtime analysis shows a clear advantage in efficiency, particularly for medium-sized and larger datasets, where our method outperformed MainSEL across various network conditions. This efficiency, combined with reduced communication overhead and the elimination of the initial setup phase, makes our method highly suitable for practical, real-world linkage scenarios.

In summary, our approach delivers significant performance improvements without compromising record linkage accuracy and demonstrates strong scalability. It offers a practical solution for real-world privacy-preserving record linkage tasks, with potential for further optimizations to support even larger datasets and more complex linkage requirements.

\section*{Acknowledgments}
This study is supported by the German Ministry of Research and Education (BMBF), project number 01ZZ2010. 

\section*{Data Availability}
The datasets generated and used during the current study are available in the PPRL repository, \hyperlink{https://github.com/mdppml/PPRL.git}{https://github.com/mdppml/PPRL.git}.

\bibliographystyle{plain}
\bibliography{reference}
\end{document}